\documentclass[twocolumn,twoside,slac]{revtex4}
\usepackage{graphicx}
\usepackage{fancyhdr}
\begin{document}

\title{Constraints on Neutrino Mixing Parameters with the SNO data}

%

\author{A. Bellerive}
\affiliation{Ottawa-Carleton Institute for Physics, Department of Physics \\
Carleton University, 1125 Colonel By Drive, Ottawa, K1S 5B6, Canada \\
E-mail: alain\_bellerive@carleton.ca}

\begin{abstract}
This paper reviews the constraints imposed on the solar
neutrino mixing parameters by data collected by the Sudbury Neutrino
Observatory (SNO). The SNO multivariate analysis is reviewed.
The global solar neutrino analysis is emphasized
in terms of matter-enhanced oscillation of two active flavors.
An outline of how SNO uses the data to 
produce oscillation contour plots and how to include 
the relevant correlations for the new 
salt data in similar oscillation analyses is summarized.
\end{abstract}

\maketitle

\thispagestyle{fancy}


%
\section{Introduction}

The deficit of detected neutrinos coming from the Sun compared with our 
expectations based on laboratory measurements, 
known as the Solar Neutrino Problem, was one of the outstanding 
problems in basic physics for over thirty years. 
It appeared inescapable that either our understanding of the energy producing 
processes in the Sun was seriously defective, 
or neutrinos, one of the fundamental particles in the Standard Model, had
important properties which had not been measured. 
It was indeed argued by some that we needed to change our ideas on how energy 
was produced in fusion reactions inside the Sun. 
Others suggested that the problem arose due to peculiar characteristics of 
neutrinos such as vacuum or matter oscillations. 
It is useful to review the evolution of our understanding from the data 
collected by various solar neutrino experiments. 
The new analysis of the salt data collected by the Sudbury Neutrino 
Observatory (SNO)~\cite{bib:snosalt} 
will be described, together with the technique used to combine the results of
many solar neutrino experiments.

\section{Solar Neutrinos}

The energy in the Sun is produced by nuclear reactions 
that transform hydrogen into helium. Through the fusion reactions,
four protons combine to form a helium nucleus containing two protons and two 
neutrons. The only reactions that allow 
this to happen are caused by weak interactions like nuclear beta decay. 
Each time a neutron is formed, there must be an associated positron 
and electron neutrino produced. Neutrinos can travel directly from the core of 
the Sun to the Earth in a about eight minutes and hence provide a direct way 
to study thermonuclear processes in the Sun. The detailed predictions of 
the solar electron neutrino flux have been produced
by John Bahcall and his collaborators from the 1960's until now. 
Their calculations are refereed to as the Standard Solar 
Model (SSM). In this proceeding, the Bahcall-Pinsonneault 
calculations~\cite{bib:BP} are compared to experimental results. 

It is known that neutrinos exist in different flavors corresponding to the three 
charged leptons: the electron, 
muon, and tau particles. If neutrinos have masses, flavor can mix 
and a neutrino emitted in a weak interaction is represented as
a superposition of mass eigenstates.
In the case of three flavors of neutrino, the mixing matrix $U$ is 
called the Maki-Nakagawa-Sakata-Pontecorvo (MNSP) matrix~\cite{bib:mnsp} and
\( \nu_\ell = \sum_i U_{\ell i} |\nu_i\rangle \).
Here the neutrino mass eigenstates are denoted by $\nu_i$ with $i = 1, 2, 3$, while 
the flavor eigenstates are labeled $(e, \mu, \tau)$. 
The most general form of mixing for
three families of neutrinos can be simplified 
so that only two neutrinos participate in the oscillations.
Hence, the survival probability for solar neutrinos propagating in 
time takes the approximate form
\begin{equation}
\label{eq:survive}
P_{e \beta} = 
\delta_{e \beta}-(2\delta_{e \beta}-1) \sin^2 2\theta\sin^2 (1.27\frac{\Delta m^2 L}{E}) \, .
\end{equation}
The mixing angle is represented by $\theta$, $L$ is the distance between the production 
point of $\nu_e$ and the point of detection of $\nu_{\beta}$,
E is the energy of the neutrino, and $\Delta m^2 \equiv m^2_j - m^2_i$
is the difference in the squares of the masses of the two states $\nu_j$ and $\nu_i$
which are mixing. The function $\delta_{e \beta}$ is the usual Kronecker delta.
The numerical constant 1.27 is valid for $L$ in meters, 
$E$ in MeV, and $\Delta m^2$ in eV$^2$. The energy
of a neutrino depends on the type of nuclear reaction
which produced it. By studying the evolution of the solar 
neutrinos as a function of $L$, all the
physics is embedded in one angle $\theta$,  one
mass difference $\Delta m^2$, and the sign of $\Delta m^2$. 
This corresponds to the extraction
of the three MNSP elements: $U_{e1}$, $U_{e2}$, and $U_{e3}$.

\section{Sudbury Neutrino Observatory}

The Sudbury Neutrino Observatory (SNO) is a 1,000 ton heavy-water
$\check{\rm{C}}$erenkov detector\cite{bib:snonim} situated 2~km underground in INCO's 
Creighton mine in Canada. Another 7,000 tons of ultra-pure light water 
is used for support and shielding. The heavy water is in 
an acrylic vessel (12~m diameter and 5~cm thick) viewed by 9,456 PMT
mounted on a geodesic structure 18~m in diameter; all contained within a 
polyurethane-coated
barrel-shaped cavity (22~m diameter by 34~m high).
The solar-neutrino detectors in operation prior to SNO were mainly sensitive 
to the electron neutrino type; while the use of heavy water by SNO allows 
neutrinos to interact through 
charged-current (CC), elastic-scattering (ES), or neutral-current (NC) 
interactions.
The determination of these reaction rates 
is a critical measurement in determining if neutrinos oscillate in transit 
between the core of the Sun and their observation on Earth.


During the pure $D_2O$ phase of the experiment, the signal was determined
with a statistical analysis based on the direction, $\cos\theta_{\rm{sun}}$,
the position, $R$, and the kinetic energy, $T$, of the reconstructed events
assuming the SSM energy spectrum shape~\cite{bib:ortiz}. The final 
selection criteria were $T \geq 5$~MeV and $R \leq 550$~cm. The result
of the extended maximum-likelihood fit yields~\cite{bib:snod2o} 
\begin{eqnarray}
\Phi_{\rm{CC}} & = & 
1.76 \, ^{+0.06} _{-0.05} \, ^{+0.09} _{-0.09} \, \times 10^6~{\rm cm}^{-2} {\rm s}^{-1} \, , \nonumber \\
\Phi_{\rm{ES}} & = & 
2.39 \, ^{+0.24} _{-0.23} \, ^{+0.12} _{-0.12} \, \times 10^6~{\rm cm}^{-2} {\rm s}^{-1} \, ,\\
\Phi_{\rm{NC}} & = & 
5.09 \, ^{+0.44} _{-0.43} \, ^{+0.46} _{-0.43} \, \times 10^6~{\rm cm}^{-2} {\rm s}^{-1} \, . \nonumber
\end{eqnarray}
The excess of the NC flux over the CC and ES fluxes implies neutrino flavor transformations.
There is also a good agreement between the SNO NC flux and the total $^8B$ flux 
of $5.05 ^{+1.01} _{-0.81}$ $\times 10^6~{\rm cm}^{-2} {\rm s}^{-1}$ predicted by the SSM.
A simple change of variables that resolves the data directly into electron
and non-electron components~\cite{bib:snod2o} indicates 
clear evidence of solar neutrino 
flavor transformation at 5.3 standard deviations
\begin{eqnarray}
\phi_{e}       & = & 1.76 \, ^{+0.06} _{-0.05} \, ^{+0.09} _{-0.09} \, \times 10^6~{\rm cm}^{-2} {\rm s}^{-1} \, ,\\
\phi_{\mu\tau} & = & 3.41 \, ^{+0.45} _{-0.45} \, ^{+0.48} _{-0.45} \, \times 10^6~{\rm cm}^{-2} {\rm s}^{-1} \, .
\end{eqnarray}

Allowing a time variation of the total flux of solar neutrinos 
leads to day/night measurements by SNO, which are sensitive to 
the neutrino type~\cite{bib:snoDN} 
\begin{eqnarray}
A_{\rm{DN}}({\rm{total}}) = (-24.2 \pm 16.1 \, ^{+2.4} _{-2.5}) \, \% \, , \\
A_{\rm{DN}}(e) = (12.8 \pm 6.2 \, ^{+1.5} _{-1.4}) \, \% \, .
\end{eqnarray}
By forcing no asymmetry in the $\phi_{e}+\phi_{\mu\tau}$ 
rate, i.e. $A_{\rm{DN}}({\rm{total}})=0$,
the day/night asymmetry for the electron neutrino is~\cite{bib:snoDN}
\( A_{\rm{DN}}(e) = (7.0 \pm 4.9 \, ^{+1.3} _{-1.2}) \).


SNO published its
first results of the salt phase~\cite{bib:snosalt} in coincidence
with the PHYSTAT2003 conference. The measurements were made 
with dissolved $NaCl$ in the heavy water
to enhance the sensitivity and signature for neutral-current interactions.  
Neutron capture on $^{35}Cl$ typically produces multiple $\gamma$ rays while 
the CC and ES reactions produce single electrons. The greater isotropy of
the $\check{\rm{C}}$erenkov light from neutron capture events relative to CC and ES events 
allows good statistical separation of the event types. 
The degree of the $\check{\rm{C}}$erenkov light isotropy is determined by the pattern 
of PMT hits. This separation allows a precise 
measurement of the NC flux to be made independent of assumptions about 
the CC and ES energy spectra. 
To minimize the possibility of introducing biases, SNO performed a blind analysis 
for the model independent determination of the total 
active $^8B$ solar neutrino. In this analysis, events are
statistically separated 
into CC, NC, ES, and external-source neutrons using an extended maximum-likelihood 
technique based on 
the distributions of isotropy, $\cos\theta_{\rm{sun}}$, and radius, R, 
within the detector. To take into account correlations between 
isotropy and energy, a 2D joint probability density function (PDF) is constructed.
This analysis differs from the analyses of the pure $D_2O$ data~\cite{bib:snod2o,bib:snoDN} 
since (1) correlations are explicitly incorporated in the signal extraction 
and (2) the spectral distributions of the ES and CC events are not constrained 
to the $^8B$ shape, but 
are extracted from the data. $\check{\rm{C}}$erenkov event backgrounds from $\beta-\gamma$ decays 
are reduced with an effective electron kinetic energy 
threshold $T$ $\geq$ 5.5~MeV and a fiducial volume
with radius $R \leq 550$ cm. 

The extended maximum-likelihood analysis 
gives the following $^8B$ fluxes~\cite{bib:snosalt}
\begin{eqnarray}
\Phi_{\rm{CC}} & = & 
1.59 \, ^{+0.08}_{-0.07} \, ^{+0.06}_{-0.08} \,\times 10^6~{\rm cm}^{-2} {\rm s}^{-1} \, , \nonumber \\
\Phi_{\rm{ES}} & = & 
2.21 \, ^{+0.31}_{-0.26} \pm{0.10} \, \times 10^6~{\rm cm}^{-2} {\rm s}^{-1} \, ,\\
\Phi_{\rm{NC}} & = & 
5.21 \pm 0.27 \pm 0.38 \, \times 10^6~{\rm cm}^{-2} {\rm s}^{-1} \, . \nonumber
\end{eqnarray}
The systematic uncertainties on the derived fluxes are shown in Table~\ref{errors}.
These fluxes are in agreement with previous SNO measurements and the SSM.  
The ratio of the $^8B$ flux measured with the CC and NC reactions then provides
confirmation of solar neutrino oscillations
\begin{equation}
\frac{\Phi_{\rm{CC}}}{\Phi_{\rm{NC}}}  =  0.306 \pm 0.026 \pm 0.024 \, . \\
\end{equation}

\begin{table}
\begin{tabular}{|llll|}
\hline
Source               & NC  & CC  & ES  \\ \hline\hline 
Energy scale         & -3.7,+3.6 & -1.0,+1.1 & $\pm1.8$ \\ 
Energy resolution    & $\pm1.2$ & $\pm0.1$ & $\pm0.3$ \\ 
Energy non-linearity & $\pm0.0$ & -0.0,+0.1 & $\pm0.0$ \\ 
 Radial accuracy     & -3.0,+3.5 & -2.6,+2.5 & -2.6,+2.9 \\ 
Vertex resolution    & $\pm0.2$ & $\pm0.0$ & $\pm0.2$ \\ 
Angular resolution   & $\pm0.2$ & $\pm0.2$ & $\pm2.4$ \\ 
Isotropy mean        & -3.4,+3.1 & -3.4,+2.6 & -0.9,+1.1 \\ 
Isotropy resolution  & $\pm0.6$ & $\pm0.4$ & $\pm0.2$ \\ 
Radial energy bias   & -2.4,+1.9 & $\pm0.7$ & -1.3,+1.2 \\ 
Vertex Z accuracy    & -0.2,+0.3 & $\pm0.1$ & $\pm0.1$ \\ 
Internal neutrons    & -1.9,+1.8 & $\pm0.0$ & $\pm0.0$ \\ 
Internal background  & $\pm0.1$ & $\pm0.1$ & $\pm0.0$ \\ 
 Neutron capture     & -2.5,+2.7 & $\pm0.0$ & $\pm0.0$ \\ 
$\check{\rm{C}}$erenkov backgrounds & -1.1,+0.0 & -1.1,+0.0 & $\pm0.0$ \\ 
AV events            & -0.4,+0.0& -0.4,+0.0 & $\pm0.0$ \\ \hline
Total uncertainty    &-7.3,+7.2 & -4.6,+3.8 &-4.3,+4.5 \\ \hline 
\end{tabular}
\caption{\label{errors}Systematic uncertainties (in \%) on fluxes for 
the spectral shape unconstrained analysis of the salt data set.}
\end{table}

\section{How to Use the SNO Data}

The SNO CC, ES and NC fluxes are statistically correlated, 
since they are derived from a fit to a single data set.
The statistical correlation coefficients 
between the fluxes in the salt phase are
\begin{eqnarray}
\rho_{{\rm{CC,NC}}} & = & -0.521 \, , \nonumber \\
\rho_{{\rm{CC,ES}}} & = & -0.156 \, , \\
\rho_{{\rm{ES,NC}}} & = & -0.064 \, . \nonumber
\end{eqnarray}
These can be used with the statistical uncertainties quoted by SNO~\cite{bib:snosalt} 
to write down the statistical covariance matrix for the salt fluxes.
Systematic uncertainties between fluxes can be correlated as well. Some sources of
systematic error, such as neutron capture efficiency, affect only one of the 
three fluxes, and so can be considered to be uncorrelated with the other fluxes. 
Other systematics can be either 100\% correlated (e.g. radial accuracy) or 
100\% anticorrelated (e.g. isotropy mean). 
The most important anti-correlated systematic is the isotropy mean. Isotropy is 
important for separating CC and ES events from NC events, so CC and ES will have 
a negative correlation with the NC flux (and a positive correlation with each other) 
for the isotropy uncertainty.
Table~\ref{tab:cor} shows the sign of the correlation for each systematic 
of Table~\ref{errors}. Using the table of systematics
and the signs for the correlations, one can assemble an individual covariance 
matrix for each systematic. Then, to get the total covariance matrix for the CC, 
ES and NC fluxes, one simply adds all of the covariance matrices together.

Even when fluxes are being analyzed as opposed to energy spectra, it is best to 
determine the effect of energy-related systematics at each grid point in 
the $\Delta m^2 - \tan^2\theta$ plane. For the salt analysis, these include energy 
scale and energy resolution; the uncertainty due to energy non-linearity is tiny 
so that it can reasonably be ignored. The energy scale uncertainty is
implemented as a 1.1\% uncertainty in the total energy; while the energy resolution 
has an uncertainty which is energy dependent for $T > 4.975$~MeV
\begin{equation}
\frac{\Delta \sigma_T}{\sigma_T} = 
0.035 + 0.00471 \times (T-4.975) \, ,
\end{equation}
and $\frac{\Delta \sigma_T}{\sigma_T}=0.034$ 
for $T < 4.975$~MeV. Here $T$ is the reconstructed kinetic 
energy. For all other systematics, it is
assumed that the effect on the fluxes is the same for all oscillation parameters.
\begin{table}
\begin{tabular}{|llll|}
\hline
Source               & NC  & CC  & ES  \\ \hline\hline 
Energy scale         & +1 & +1 & +1 \\ 
Energy resolution    & +1 & +1 & +1 \\ 
Energy non-linearity & +1 & +1 & +1 \\ 
Radial accuracy      & +1 & +1 & +1 \\ 
Vertex resolution    & +1 & +1 & +1 \\ 
Angular resolution   & +1 & +1 & \,--\,1 \\ 
Isotropy mean        & +1 & \,--\,1 & \,--\,1 \\ 
Isotropy resolution  & +1 & +1 & +1 \\ 
Radial energy bias   & +1 & +1 & +1 \\ 
Vertex X accuracy    & +1 & +1 & +1 \\ 
Vertex Y accuracy    & +1 & +1 & +1 \\ 
Vertex Z accuracy    & +1 & \,--\,1 & \,--\,1 \\ 
Internal neutrons    & +1 & ~~0 & ~~0 \\ 
Internal background  & +1 & +1 & +1 \\
Neutron capture      & +1 & ~~0 & ~~0 \\
$\check{\rm{C}}$erenkov backgrounds & +1 & +1 & +1 \\
AV events            & +1 & +1 & +1 \\
\hline
\end{tabular}
\caption{\label{tab:cor} Signs of systematic correlations, relative to its effect 
on the NC flux. An entry of $+1$ indicates a 100\% positive correlation, $-1$ a 
100\% negative correlation, and $0$ means no correlation.}
\end{table}

When SNO quotes $\Phi_{\rm{CC}} = 1.59 \times 10^6~{\rm cm}^{-2} {\rm s}^{-1}$, 
it refers to the integral flux from zero to the endpoint assuming an undistorted
$^8B$ spectrum. It implies
that the number of events attributed to CC interactions above $T=5.5$~MeV is equal to 
the number of events that would be observed if the $\nu_e$ flux follows 
the $^8B$ spectral shape. The $^8B$ spectral shape aspect of this definition is only 
for normalization; there is no assumption of any spectral shape 
when extracting the number of events during the salt phase. Similar definitions 
apply for the NC and ES fluxes.

For the comparison of the SNO CC rate with the theoretical rates for a set of oscillation 
parameters, the  $\Phi_{\rm{CC}}$ flux is
\begin{equation}
f_B \int_0^\infty \phi_{\rm{SSM}}(E_\nu) \, dE_\nu \, S(T,T_e,E_\nu) \, , 
\end{equation}
with the scale $S(T,T_e,E_\nu)$ is equal to
\begin{equation}
\frac{\int_0^\infty \int_0^\infty \int_{5.5}^\infty F(T,T_e,E_\nu) P_{ee}(E_\nu) dT dT_e dE_\nu}
{\int_0^\infty \int_0^\infty \int_{5.5}^\infty F(T,T_e,E_\nu) dT dT_e dE_\nu} \, ,
\end{equation}
where
\begin{equation}
F(T,T_e,E_\nu) = \phi_{\rm{SSM}}(E_\nu) \, \frac{d\sigma(E_\nu,T_e)}{dT_e} \, N(T_e,\sigma^2_T) \, .
\end{equation}
The factor $f_B$ allows the total $^8B$ solar neutrino flux to float from the SSM value,
$E_\nu$ is the neutrino energy, $P_{ee}$ is the survival probability, 
$T_e$ is the true recoil electron kinetic energy, and
$T$ is the observed electron kinetic energy; while 
$N(T_e,\sigma^2_T)$ is a Gaussian energy response function 
for $T$ with 
\(
\sigma_T(T) = -0.145 + 0.392 \sqrt{T} + 0.0353 T \, .
\)
It is a similar definition for the SNO ES flux, remembering to include the contribution 
from $\nu_{\mu\tau}$ using the appropriate cross section and $(1-P_{ee})$. There is 
no ambiguity in interpreting NC flux since it is equal to the total SSM flux.

\section{Global Fits}

This section summarizes the constraints from solar neutrino data in a global analysis.
The allowed region in the oscillation $\Delta m^2 - \tan^2\theta$ plane is obtained by 
comparing the measured rates to the calculated SSM solar neutrino 
rate.
We consider a set of $N$ observables $R_n$ for $n=1,2, \cdots ,N$ with the 
associated set of experimental observations $R_n^{\rm{exp}}$ and theoretical 
predictions $R_n^{\rm{th}}$. In general, one wants to build a $\chi^2$ function which 
measures the differences $R_n^{\rm{exp}}-R_n^{\rm{th}}$ in units of the total experimental 
and theoretical uncertainties. This task is completely determined from the
estimated uncorrelated errors $u_n$ and a set of
correlated systematic errors $c_n^k$ caused by $K$ independent sources. The
correlation coefficients between the different observables are 
$\rho(u_n,u_m)=\pm\delta_{nm}$ and $\rho(c_n^k,c_m^h)=\pm\delta_{kh}$.
The covariance matrix takes the form
\(
\sigma^2_{nm} = \delta_{nm}u_n u_m + \sum_{k=1}^K c_n^k c_m^h 
\)
and all the experimental information is combined together in a global $\chi^2$
\begin{equation}
\chi^2_{\rm{cov}} = \sum_{n,m=1}^N (R_n^{\rm{exp}} - R_n^{\rm{th}}) [\sigma^2_{nm}]^{-1}
(R_m^{\rm{exp}} - R_m^{\rm{th}}) \, .
\end{equation}

\begin{figure}
\includegraphics[width=88mm]{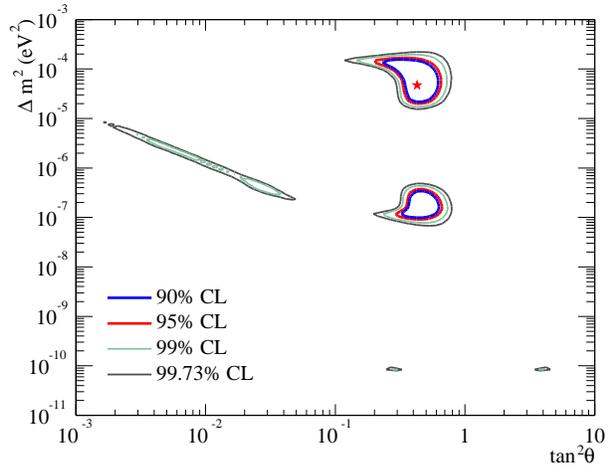}
\caption{ SNO-only neutrino oscillation contours, including pure $D_2O$ day/night spectra, 
salt~CC, NC, ES fluxes, with $^8B$ flux free and {\emph{hep}} flux fixed.
The best-fit point is $\Delta m^{2}=4.7\times10^{-5}$, $\tan^{2}\theta=0.43$, $f_{B}=1.03$, 
with $\chi^{2}$/d.o.f.=26.2/34. The inside of the covariance regions is allowed.}
\label{fig:sno}
\end{figure}

The salt shape-unconstrained fluxes presented here, combined with shape-constrained fluxes
and day/night energy spectra from the pure $D_2O$ phase~\cite{bib:snod2o,bib:snoDN}, place 
impressive constraints on the allowed neutrino flavor mixing parameters. In the fit,
the ratio $f_{B}$ of the total $^8B$ flux to the SSM value is a free parameter
together with the mixing parameters. A combined $\chi^2$ fit to SNO $D_2O$ and salt 
data alone yields the allowed regions in
$\Delta m^2$ and $\tan^2 \theta$ shown in Fig.~\ref{fig:sno}.  
There are certainly correlations between the salt and the $D_2O$ phase, since it's 
the same detector. However, these 
correlations are estimated to be negligibly small.

The $\chi^2_{\rm{cov}}$ calculated above from the SNO NC, CC and ES fluxes 
is added to a global 
analysis which includes data from all the other solar neutrino experiments. Systematic 
errors that are correlated between different experiments, such as cross section 
uncertainties or uncertainties on the $^8B$, are accounted for by including the 
covariance terms between different experimental results. The effect of 
the $^8B$ spectral shape uncertainty is determined at each grid 
point in the oscillation plane. 

\begin{figure}
\includegraphics[width=88mm]{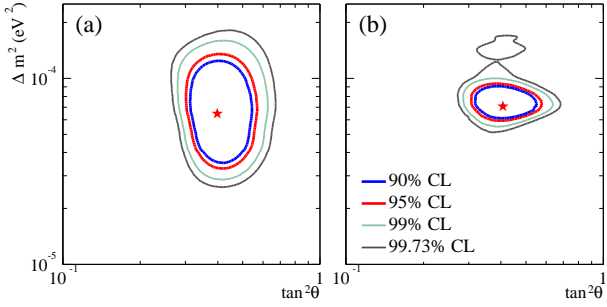}
\caption{Allowed region of the $\Delta m^2 - \tan^2\theta$ plane 
determined by a $\chi^2$ fit to (a) 
the Chlorine, Gallium, SK, and SNO experiments. The best-fit point 
is $\Delta m^2=6.5\times10^{-5}$, $\tan^{2}\theta=0.40$, $f_{B}=1.04$, 
with $\chi^{2}$/d.o.f.=70.2/81. (b) Solar global + KamLAND.
The best-fit point 
is $\Delta m^2=7.1\times10^{-5}$, $\tan^{2}\theta=0.41$, $f_{B} = 1.02$.  
The inside of the covariance contours is the allowed region.}
\label{fig:global}
\end{figure}

The global analysis includes the Homestake results~\cite{bib:cl}, the updated Gallium
flux measurements~\cite{bib:lownu,bib:nu2002}, the SK zenith 
spectra~\cite{bib:sktime}, and the $D_2O$ and salt results 
from SNO~\cite{bib:snod2o,bib:snoDN,bib:snosalt}. 
At each grid point in the $\Delta m^2 - \tan^2\theta$ plane, the expected
rate for each energy bin is calculated and compared to
the measured rate.
The free parameters in the global fit are the total $^8B$ flux,
the difference of the squared masses $\Delta m^2$, and the mixing 
angle $\theta$. The higher energy $hep$ $\nu_e$ flux is fixed 
at $9.3 \times 10^3$~cm$^{-2}$~s$^{-1}$. Contours are 
generated in $\Delta m^2$ and $\tan^2\theta$
for $\Delta \chi^2_{\rm{cov}}$ = 4.61 (90\% CL), 5.99 (95\% CL), 9.21 (99\% CL), 
and 11.83 (99.73\% CL). 
We assume a Gaussian distribution of $R_n^{\rm{exp}}$
for a given value of the true parameters
$\delta m^2$ and $\tan^2\theta$ when we map the survival probability 
into the MSW plane~\cite{bib:freq}. As presented in Fig~\ref{fig:global}(a),
the combined results of all solar neutrino experiments can be used to
determine a unique region of the oscillation parameters; 
the allowed region in this parameter space shrinks considerably 
to a portion of the Large Mixing Angle (LMA) region.

A global analysis including the KamLAND reactor anti-neutrino 
results~\cite{bib:kamland} shrinks the allowed region further, with a best-fit 
point of $\Delta m^{2} = 7.1^{+1.2}_{-0.6}\times10^{-5}$~eV$^2$ and
$\theta = 32.5^{+2.4}_{-2.3}$ degrees, where the errors reflect $1\sigma$ constraints on 
the 2-dimensional region. This is summarized in Fig.~\ref{fig:global}(b).
With the new SNO measurements, the allowed region is constrained to only the lower band 
of LMA at $>99\%$ CL.
The best-fit point with a one dimensional projection of the uncertainties in 
the individual parameters (marginalized uncertainties) 
is $\Delta m^{2} = 7.1^{+1.0}_{-0.3}\times10^{-5}$~eV$^2$ 
and $\theta = 32.5^{+1.7}_{-1.6}$ degrees. This disfavors maximal mixing at a 
confidence level equivalent 
to 5.4 standard deviations and indicates $\tan^2\theta<1$. In our interpretation, 
the $\chi^2_{\rm{cov}}$ for $\theta = 45.0$ is $5.4^2$ higher than the best LMA fit.
The solution $\tan^2\theta<1$ corresponds to the neutrino mass 
hierarchy $m_2 > m_1$.

\section{Pull Analysis}

The pull method allows a split of the residuals from the observables and the systematic
uncertainties~\cite{bib:fogli}. This alternative approach embeds the effect of 
each independent $k^{\rm{th}}$ source of systematics through a shift of the difference 
$(R^{\rm{exp}}_n - R^{\rm{th}}_n)$ by an amount $\epsilon_k c_n^k$.
The normalization condition for the $K$ independent sources
of systematic uncertainty is implemented through quadratic 
penalties in the global $\chi^2$, which is 
minimized with respect to all $\epsilon_k$'s
\begin{equation}
\chi^2_{\rm{pull}} = \sum_{n=1}^N 
\left(\frac{R_n^{\rm{exp}} - R_n^{\rm{th}}-\sum_{k=1}^K \epsilon_k c_n^k} {u_n}\right)^2
+ \sum_{k=1}^K \epsilon_k^2 \, .
\end{equation}
In an experimental context, the pull approach is not blind since it uses the data to
constrain the systematic uncertainties. Systematic shifts 
calculated with the pull method should not be used as iterative corrections to
experimental systematic uncertainties since it might lead to biases in the estimation
of the mixing parameters. Nevertheless, the pull 
approach provides a nice framework to study each component of a global fit after
a detailed study of the systematic uncertainty of each observables. See details 
in Ref.~\cite{bib:fogli}.

\section{Summary}

A summary of how to use the new salt data published by SNO is described in the 
context of solar neutrino analyses
of matter-enhanced oscillation of two active flavors.
Solar neutrino oscillation is clearly established by SNO.
Matter effects~\cite{bib:msw} explain the energy dependence of solar 
oscillations with Large Mixing Angle (LMA) solutions favored.
The global analysis of the solar and reactor neutrino results yields 
$\Delta m^{2} = 7.1^{+1.0}_{-0.3}\times10^{-5}$~eV$^2$ and 
$\theta = 32.5^{+1.7}_{-1.6}$ degrees.  

SNO is presently analyzing its full salt data set with a detailed treatment
of the day/night and spectral information. In the future SNO will
perform a global oscillation fit with a maximum-likelihood method.


\begin{acknowledgments}
This article builds upon the careful and detailed work of
many people.
Special thanks for the contributions of M. Boulay, M. Chen, S. Oser,
Y. Takeuchi, G.~Te\v{s}i\'{c}, and D. Waller.
This research has been financially supported in
Canada by the Natural Sciences and Engineering Research 
Council (NSERC), the Canada Research Chair (CRC) Program,
and the Canadian Foundation for Innovation (CFI).
\end{acknowledgments}


\end{document}